\documentclass[aps,pra,twocolumn,amsmath,amssymb,floatfix,superscriptaddress]{revtex4-2}
\usepackage{amsmath}
\usepackage{amssymb}
\usepackage{amsfonts}
\usepackage{bbm}
\usepackage{listings}
\usepackage{enumerate}
\usepackage{latexsym}
\usepackage{multirow}
\usepackage[T1]{fontenc}
\usepackage{graphicx}
\usepackage{braket}
\usepackage[pdftex,colorlinks=false]{hyperref}
\usepackage{xcolor, colortbl}
\usepackage{verbatim}

\usepackage{times}
\usepackage{array, makecell}
\usepackage[export]{adjustbox}
\usepackage{tabularx}
\usepackage{mathtools}
\usepackage{tikz}
\usepackage{siunitx}

\usepackage{lineno}

\newcolumntype{M}[1]{>{\centering\arraybackslash}m{#1}} 

\newcolumntype{N}{@{}m{0pt}@{}}

\newcommand{\beq}{\begin{equation}}
\newcommand{\eneq}{\end{equation}}

\def\be{\begin{equation}}
\def\ee{\end{equation}}
\def\ba{\begin{eqnarray}}
\def\ea{\end{eqnarray}}

\def\R{{\rm Re}}
\def\Z{\mathbb{Z}}
\def\C{\mathbb{C}}

\def\dag{\dagger}

\def\beq{\begin{equation}}
\def\eeq{\end{equation}}
\def\barray{\begin{eqnarray}}
\def\earray{\end{eqnarray}}

\font\upright=cmu10 scaled\magstep1
\def\stroke{\vrule height8pt width0.4pt depth-0.1pt}

\def\Zmath{\mathbb{Z}}
\def\Qmath{\vcenter{\hbox{\upright\rlap{\rlap{Q}\kern
                   3.8pt\stroke}\phantom{Q}}}}
\def\Nmath{\vcenter{\hbox{\upright\rlap{I}\kern 1.7pt N}}}
\def\Cmath{\vcenter{\hbox{\upright\rlap{\rlap{C}\kern
                   3.8pt\stroke}\phantom{C}}}}
\def\Rmath{\vcenter{\hbox{\upright\rlap{I}\kern 1.7pt R}}}
\def\Z{\ifmmode\Zmath\else$\Zmath$\fi}
\def\Q{\ifmmode\Qmath\else$\Qmath$\fi}
\def\N{\ifmmode\Nmath\else$\Nmath$\fi}
\def\C{\ifmmode\Cmath\else$\Cmath$\fi}
\def\R{\ifmmode\Rmath\else$\Rmath$\fi}

\input{epsf}

\newcounter{defcounter}
\newcommand{\affA}{Van der Waals-Zeeman Institute, Institute of Physics,
University of Amsterdam, 1098 XH Amsterdam, Netherlands}
\newcommand{\affB}{QuSoft, Science Park 123, 1098 XG Amsterdam, the Netherlands}
\newcommand{\affC}{Institute for Theoretical Physics, Institute of Physics, University of Amsterdam, Science Park 904, 1098 XH Amsterdam, the Netherlands}

\begin{document}

%
%
%

\title{Trapped Ion Quantum Computing using Optical Tweezers and Electric Fields}

\author{M. Mazzanti}\affiliation{\affA}
\author{R. X. Sch{\"u}ssler}\affiliation{\affA}
\author{J. D. Arias Espinoza}\affiliation{\affA}
\author{Z. Wu}\affiliation{\affA}
\author{R. Gerritsma}\affiliation{\affA}\affiliation{\affB}
\author{A. Safavi-Naini}\affiliation{\affB}\affiliation{\affC}

\begin{abstract}
We propose a new scalable architecture for trapped ion quantum computing that combines optical tweezers delivering qubit state-dependent local potentials with oscillating electric fields. Since the electric field allows for long-range qubit-qubit interactions mediated by the center-of-mass motion of the ion crystal alone, it is inherently scalable to large ion crystals. Furthermore, our proposed scheme does not rely on either ground state cooling or the Lamb-Dicke approximation. We study the effects of imperfect cooling of the ion crystal, as well as the role of unwanted qubit-motion entanglement, and discuss the prospects of implementing the state-dependent tweezers in the laboratory. 
\end{abstract}

\date{\today}

\maketitle

\noindent \emph{Introduction.} Trapped ions form one of the most mature laboratory systems for quantum information processing and quantum simulation~\cite{Cirac:1995,Porras:2004,Blatt:2008}. Many of the basic building blocks needed for these technologies have been demonstrated: high fidelity detection and preparation~\cite{Myerson:2008}, and universal quantum operations performed by external fields coupling to the internal states of the ions. While quantum gates have been performed with very high fidelities in trapped ions~\cite{Brown:2011,Ballance:2016,Gaebler:2016}, scaling up the system while maintaining the quality of operations has proven to be challenging. In particular, as the length of ion crystals increases, the number of motional modes to which the gate lasers couple also increases. This leads to a reduction of interaction strength for gates between distant qubits~\cite{Kim:2009}. Furthermore, the number of degrees of freedom with which the qubits can erroneously entangle increases.

In this work we propose a novel universal trapped ion quantum computing architecture which uses state-dependent optical tweezer potentials~\cite{Olsacher:2020,Teoh:2021,Espinoza:2021} combined with oscillating electric fields. This setup allows us to overcome the obstacles described above. Since the electric fields only couple to the center-of-mass (COM) mode of the ion crystal, adverse effects of spectator modes that reduce the range of interaction are avoided. Moreover, our gate does not rely on the Lamb-Dicke approximation which requires the wavepackets of the ions to be confined to a space smaller than the wavelength of the laser implementing the gate. This extends the parameter regime in which the gate can be operated. Combining the proposed two-qubit gates with single-qubit gates that can be straightforwardly delivered by tweezers, the setup can be used as a universal quantum computer. 

\begin{figure}[b]
\centering
\includegraphics[width=\linewidth]{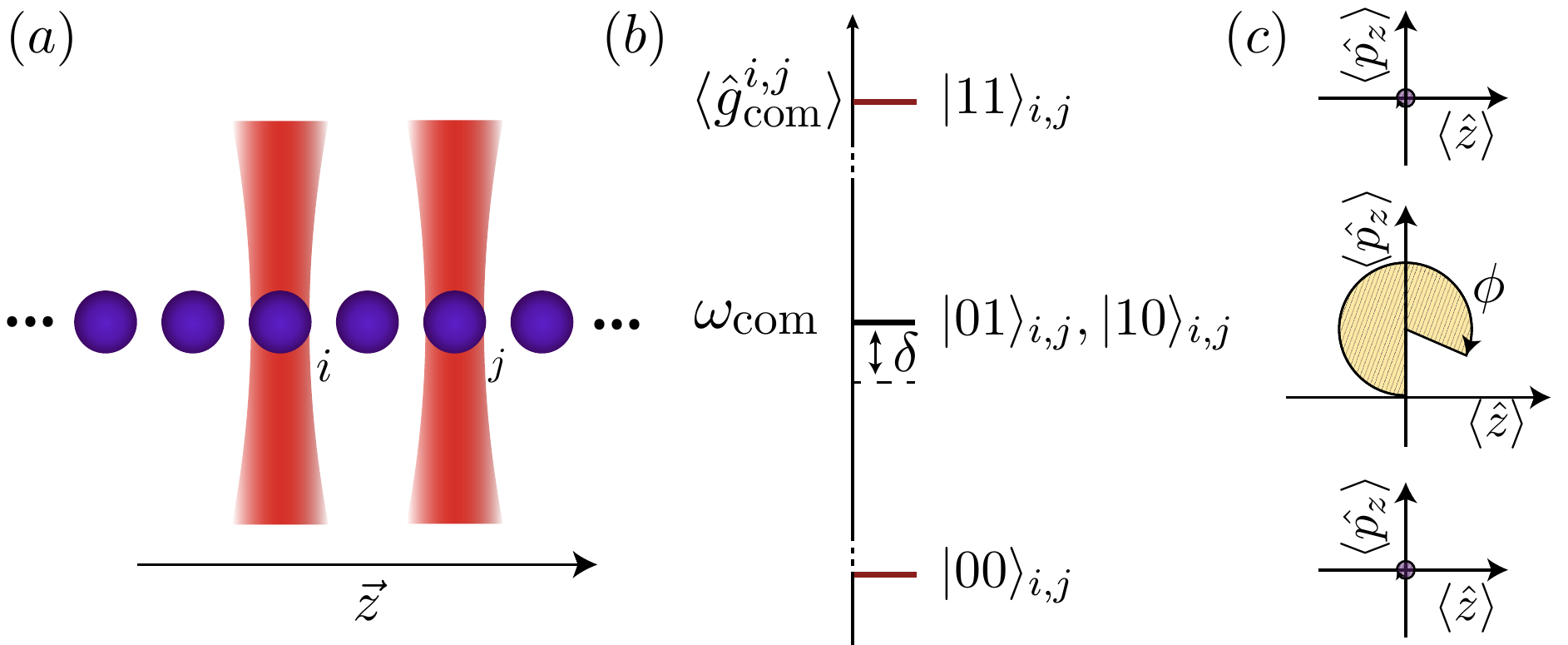}
\caption{(a) Schematic representation of a linear chain of ions where optical tweezers are applied to ions $i$ and $j$. The tweezers shift center of mass mode depending on the internal state of the pair. (b) Level scheme of the four states; only the states $|01\rangle$ and $|10\rangle$ are unaffected by the extra trapping potential generated by the tweezers. (c) Phase space dynamics of the four states when adding an electric field at a frequency $\omega_{\text{com}}-\delta$. Due to the displacement generated by the driving electric field the states $|01\rangle$ and $|10\rangle$ acquire a phase $\phi$.}
\label{fig:FigOne}
\end{figure}

We illustrate the gate mechanism applied to qubits $i$ and $j$ in Fig.~\ref{fig:FigOne}. We simultaneously apply an electric field of amplitude $E_0$ oscillating close to the COM frequency (at detuning $\delta$) and optical tweezers to the two addressed qubits. The gate mechanism works as follows: the tweezers shift the frequency of the COM mode in a state-dependent manner, so that for two qubits in the same state the electric field can no longer excite motion. In this regime, the evolution of the system is dominated by phonon mediated effective spin-spin interactions $\propto E_0^2/\delta$. We are then able to perform a geometric phase gate by choosing the appropriate electric field amplitude and detuning. Since the interactions are merely mediated by the COM mode they are independent of distance. Additionally, the required tweezer power scales linearly with the number of ions in the crystal. Both those factors contribute to the scalability of our proposal.

\emph{Realizing a geometric phase gate. }Consider a crystal of $N$ ions with masses $M$ and charge $e$ in a harmonic trap. 
The normal modes (phonon modes) and mode frequencies of the crystal can be found by diagonalizing the Hessian matrix $\mathbf A$~\cite{James:1998}. Here $A^{(ij)}=d^2V/(d\alpha_i d\alpha_j)$, where $\alpha_i$ are small deviations about the equilibrium positions of the ions, and $V$ is the total potential energy. 
The eigenvectors of the Hessian, denoted by $\mathbf b_m$, are the normal modes of the crystal. The mode frequencies are given by $\omega_m=\sqrt{\lambda_m}$ with $\lambda_m$ the eigenvalues of $\mathbf A$. 
For the 1D ion crystal considered here, the eigenmodes separate in three subclasses, corresponding to the directions of motion $x,y,z$ and in the remainder of the paper we focus on the axial direction ($z$) characterized by the trapfrequency $\omega_z$.

We address the ions of interest using tweezers formed by focused beams aligned on the equilibrium positions of the ions which leaves the geometry of the crystal independent of the qubit states. In the following, we will show that the laser parameters and qubit states can be chosen such that the dynamical polarizability of the qubit states $|0\rangle$ and $|1\rangle$ are of equal magnitude, but opposite sign. At the center of the tweezer the potential can be approximated to second order, leading to a state-dependent harmonic potential:
$\hat{H}_\text{tw}^{i,j}=\frac{1}{2} M\omega_\text{tw}^2\left(\hat{z}_i^2\hat \sigma_z^i+\hat{z}_j^2\hat \sigma_z^j\right)$. 
Here, $\hat \sigma_z^i$ ($\hat \sigma_z^j$) is the Pauli matrix operating on ion $i$ ($j$) and $\hat{z}_i$ ($\hat{z}_j$) is the position operator relative to the equilibrium position of ion $i$ ($j$). The proposed gate requires the simultaneous application of the tweezers and an oscillating electric field generated by applying an rf-voltage to an electrode close to the ion crystal. The total Hamiltonian is then given by:
\begin{equation}\label{Eq_H_tot}
    \hat{H}=\sum_m\omega_m\left(\hat{a}_m^{\dag}\hat{a}_m+\frac{1}{2}\right)+\hat{H}_\text{tw}^{i,j}+\hat{H}_\text{E}(t),
\end{equation}
\noindent with $\hat{H}_\text{E}(t)$ denoting the electric field interaction and $\hat{a}_m^{\dag}$ ($\hat{a}_m$) the creation (annihilation) operator of mode $m$.

In the limit $\omega_\text{tw} \ll \omega_m$ for all modes, we  can use perturbation theory to find the frequencies of the phonon modes in the presence of the tweezers: $\tilde{\lambda}_{m}\approx\lambda_m+\sum_{k}b_{mk}\hat{A}^{(ij)}_\text{tw}b_{mk}+...$ with $k = 1,\dots, N$. Here, the perturbation of the tweezers to the Hessian matrix is given by $\hat{A}^{(ij)}_\text{tw}=\omega_{\text{tw}}^2\left(\hat{\sigma}_z^i+\hat{\sigma}_z^j\right)$. To first order, the new mode frequencies are given by:
\begin{equation}
 \tilde{\omega}_{m}^{i,j}  \approx \sqrt{\omega_m^2+\omega_\text{tw}^2\left(b^2_{mi}\hat\sigma_z^{i}+b^2_{mj}\hat \sigma_z^{j}\right)},
\end{equation}
\noindent 
which shows that the mode frequencies shift depending on the states of qubits $i$ and $j$. 

A homogeneous electric field $E_0$ only couples to the COM motion and the resulting total Hamiltonian is, 
$$
\hat{H}_\text{E}(t)= 2\gamma (\hat{a}^{\dag}_\text{com}+\hat{a}_\text{com})\cos(\mu t), 
$$
where $\gamma=e E_0 l_\text{com}/2$, $l_\text{com} = (2M\omega_\text{com})^{-1/2}$ and  $\mu=\omega_\text{com}+\delta$ is the frequency of the electric field. The dynamics generated by the above Hamiltonian can be intuitively understood following the application of the unitary transform $\hat U_1=\exp\left[i(\delta\hat{a}_\text{com}^{\dag}\hat{a}_\text{com}+ \sum_m\omega_m\hat{a}_m^{\dag}\hat{a}_m)t\right]$, as well as the rotating wave approximation in which we neglect terms oscillating faster than $\delta$. Next, we apply a unitary transformation of the Lang-Firsov type~\cite{Lang:1968}, $\hat U_2=\exp\left[\hat{V}\left(\hat{a}^{\dag}_\text{com}-\hat{a}_\text{com}\right)\right]$, with $\hat{V}=\gamma (\hat{g}^{i,j}_\text{com}-\delta\mathbbm{1})^{-1}$. This eliminates the first order phonon coupling to arrive at the Hamiltonian:
\begin{align}
    \hat{H}_2=&\sum_m \hat{g}^{i,j}_m(\hat{a}^{\dag}_m\hat{a}_m+1/2)-\delta\hat{a}_\text{com}^{\dag}\hat{a}_\text{com}\nonumber\\
    &-\frac{\gamma^2}{2\delta}\hat \sigma_z^{i}\hat \sigma_z^{j}+\frac{\gamma^2}{g^+_\text{com}-\delta}\hat{W}_{+}+\frac{\gamma^2}{g^-_\text{com}-\delta}\hat{W}_{-},
\label{Eq_finalH}\end{align}
with $\hat{W}_{+}=|11\rangle_{ij}\langle 11|_{ij}$ and $\hat{W}_{-}=|00\rangle_{ij}\langle 00|_{ij}$ where we have dropped energy offset terms $\propto\mathbbm{1}$. The operator $\hat{g}^{i,j}_m=\tilde{\omega}_{m}^{i,j}-\omega_m$ contains the qubit state dependence. For the COM mode, $\mathbf{b}^2_{\text{com},i}=\mathbf{b}^2_{\text{com},j}=1/N$~\cite{James:1998}, resulting in $\hat{g}^{ij}_\text{com}= \sqrt{\omega_{\rm com}^2+\omega_\text{tw}^2\left(\hat{\sigma}_z^{(i)}+\hat{\sigma}_z^{(j)}\right)/N}-\omega_{\rm com}$ to the first order as schematically shown in Fig.~\ref{fig:FigOne}. Expanding the square root for $\omega_\text{tw}\ll \omega_\text{com}$, we obtain $\hat{g}^{ij}_\text{com}\approx \omega_\text{tw}^2\left(\hat{\sigma}_z^{(i)}+\hat{\sigma}_z^{(j)}\right)/(2N\omega_\text{com})$. From this it follows that the required tweezer intensity scales linearly with the number of ions in the crystal $N$. The values of $g^{\pm}_\text{com}$ are calculated by setting $ \hat{\sigma}_z^{i}+\hat{\sigma}_z^{j}\rightarrow\pm 2$. 



\emph{Effective Hamiltonian. } The first line of the Hamiltonian~\ref{Eq_finalH} contains the qubit-state dependence of the phonon modes. These may lead to residual qubit-phonon entanglement at the end of the gate which can cause errors in the gate. However, straightforward spin echo sequences can be used to undo these errors. The second line contains the qubit-qubit interactions, and the first term dominates for $|\delta| \ll |g^{\pm}|$. This can be used to implement a quantum gate that is equivalent to a geometric phase gate when setting a gate time of $\tau =2\pi/\delta$ and $\gamma^2/\delta^2=\pi/4$~\cite{Sorensen:1999,Leibfried:2003a}.


\begin{figure}[h]
\centering
\includegraphics[width=\columnwidth]{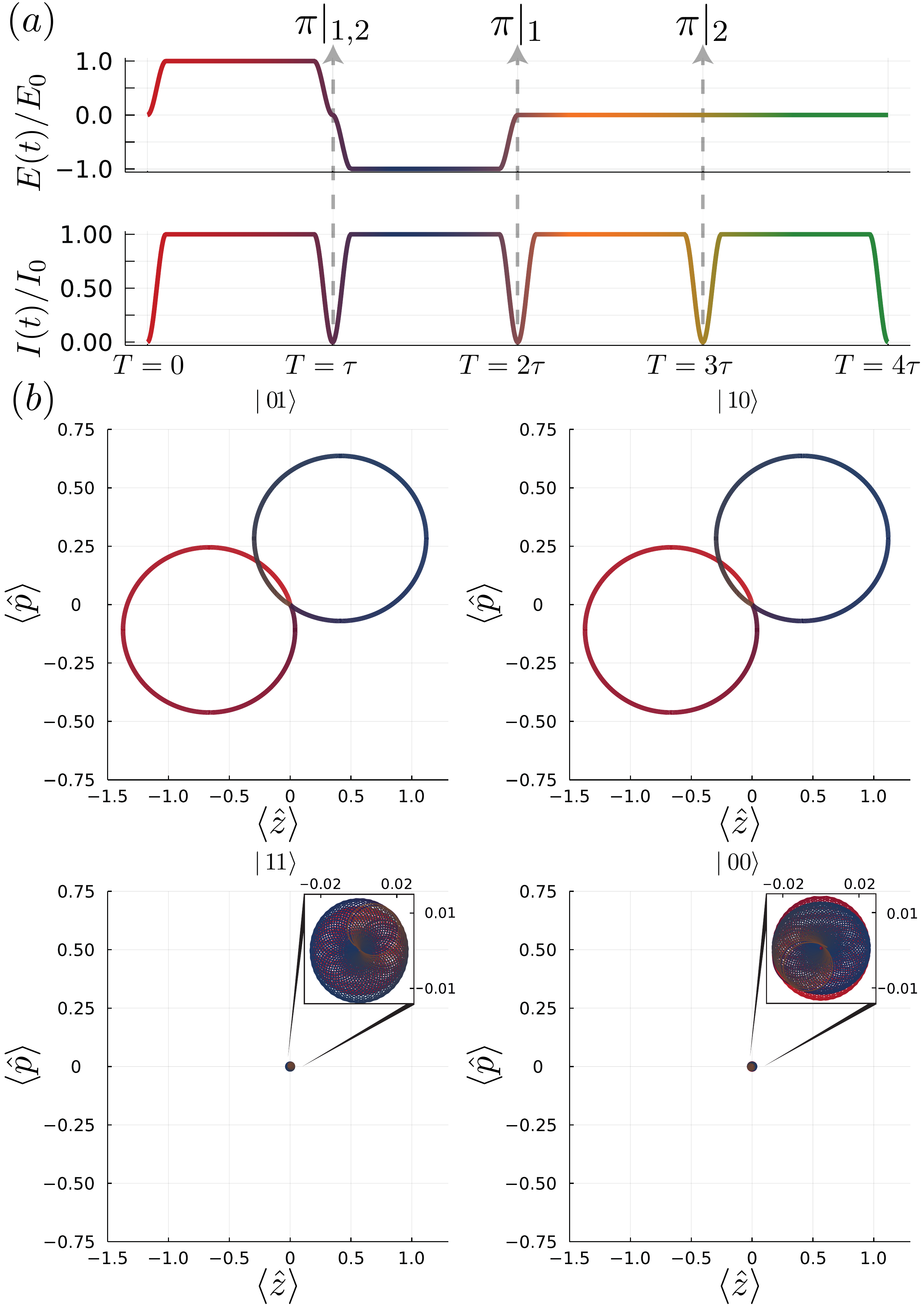}
\caption{(a) Pulse sequence used for the simulations, $E(t)/E_0$ and $I(t)/I_0$ are the normalized electric field and laser intensities, respectively. At the end of each of the first three pulses we perform a $\pi$-pulse on either ion 1 and 2, 1 or 2. (b) Resulting phase space dynamics for a two $^{171}$Yb$^+$-ion crystal in natural units. When the ions are in the state $|01\rangle$ or $|10\rangle$ they start to oscillate following the electric field. For the states $|11\rangle$ and $|00\rangle$, instead, the electric field is not resonant with the COM mode due to the shift created by the tweezers. The small motion of states $|11\rangle$ and $|00\rangle$ is highlighted in the insets. The parameters used for the simulation are: $\delta = 2\pi \times 0.001$\,MHz, electric field magnitude of $E_0 = 0.269$\,mV/m, $\omega_\text{com} = 2\pi\times1 $\,MHz and $\omega_\text{tw} = 2\pi\times 250 $\,kHz.}
\label{fig:PS_figure}
\end{figure}

In order to characterize the performance of the gate under experimental conditions, we first consider a crystal of two $^{171}$Yb$^+$ ions with trap frequency $\omega_\text{com} = 2\pi\times 1$\,MHz, and then extend our study to a crystal with $N=4$ ions to demonstrate the scalability of our scheme. We assume that the ions are initialized in a thermal state with $\bar{n}$ motional quanta. The gate sequence consists of four pulses of duration $\tau =2\pi/\delta$, as illustrated in Fig.~\ref{fig:PS_figure} (a). Each pulse uses adiabatic ramping for the electric field and laser interaction to avoid non-adiabatic coupling of phonon modes. At the end of the first pulse we apply $\pi$-pulses on both ions to remove the extra phases accumulated due to the last two terms in Eq.~\ref{Eq_finalH}. 
However, this spin echo pulse does not fully correct for the residual qubit-motion entanglement because $g_m^{i,j}(|11\rangle)+g_m^{i,j}(|00\rangle)\neq g_m^{i,j}(|01\rangle)+g_m^{i,j}(|10\rangle)$. This can be compensated with one more spin echo pulse on each ion separately. To this end, the 2nd pulse is applied to qubit 1 and the 3rd pulse to qubit 2 or vice versa,  with the electric field switched off. The complete pulse sequence can be seen in Fig.~\ref{fig:PS_figure} (a). 

\emph{Gate fidelity and scalability.} We simulate the gate dynamics generated by Eq.~\ref{Eq_H_tot} numerically and use process fidelity to characterize its performance. For the sake of simplicity, we first ignore the contribution of the stretch mode, simulating the dynamics when considering only the COM mode ($m=\text{com}$).


In Fig.~\ref{fig:PS_figure} (b) we illustrate the gate mechanism using the phase space dynamics of the four basis states for a two-ion crystal of $^{171}$Yb$^+$ prepared in the ground state of motion. For the states $\vert 01\rangle$ and $\vert 10\rangle$,  $\hat{g}^{ij}_\text{com}\approx 0$. Thus, these states follow the displacement generated by the electric field. 
On the other hand, as discussed earlier, the other two states $|11\rangle$ and $|00\rangle$ are not significantly displaced in phase space since the COM mode frequency is shifted by the tweezers. The gate parameters are set such that the phases accumulated for these four states correspond to a geometric phase gate.

For ions initialized in a thermal state with an average Fock state (FS) population $\bar{n}$, the process fidelity is given by \cite{Nielsenprocessfidelity}:
\begin{equation}
\bar{F}(\hat{U}_{\text{id}},\hat{U}_{\text{H}})=\frac{\sum_l \text{tr}\left[\hat{U}_{\text{id}}\hat{\sigma}_l^\dagger\hat{U}_{\text{id}}^\dagger\boldsymbol{\hat{\sigma}}_l(\hat{U}_{\text{H}})\right]+d^2}{d^2\left(d+1\right)},
\end{equation}
where $\boldsymbol{\hat{\sigma}}_l(\hat{U}_{\text{H}}) \equiv \text{tr}_{\text{FS}}(\hat{U}_{\text{tw}}\left[|n\rangle\langle n|\bigotimes\hat{\sigma}_l\right]\hat{U}^\dagger_{\text{H}})$ is the projector on one of the $SU(2)$ $d$-dimensional representation of Pauli matrices
(here $d$ = 4 for a two ions case) and on the Fock state $|n\rangle$, $\hat{U}_{\text{id}}$ is the ideal phase gate, and $\hat{U}_{\text{H}}$ is the unitary generated by the Hamiltonian shown in Eq.~\ref{Eq_H_tot} in the interaction picture.

\begin{figure}[h]
\centering
\includegraphics[width=\linewidth]{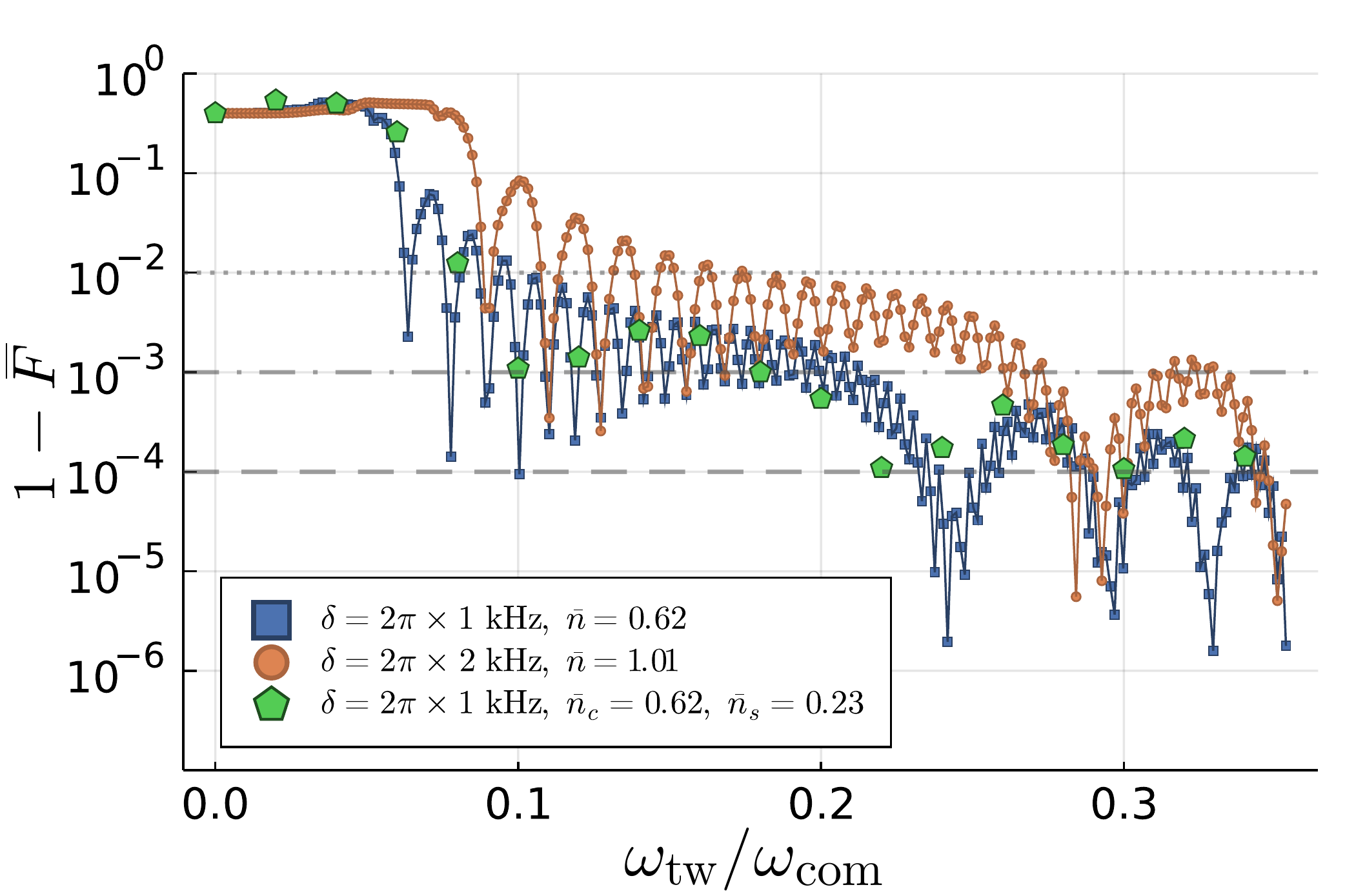}
\caption{Process fidelity as a function of tweezers strength for a two ion crystal at different $\delta$ values and thermal states $\bar{n}\simeq1$ and $\bar{n}\simeq0.6$. The given values are obtained by setting $\omega_{\text{com}} = 2\pi \times 1$\,MHz, $\delta = 2\pi \times 1 $\,kHz (blue squares) and $\delta = 2\pi \times 2 $\,kHz (orange circles), leading to a gate time of, respectively, 4\,ms and 2\,ms. The green pentagons show the fidelity for a two ion crystal taken into account the contribution of both modes. The Fock state cut-off for the thermal state used in the calculations is $n_{\text{max}}=20$ for the single mode case and $n_{\text{c},\text{max}}=14$ and $n_{\text{s},\text{max}}=10$ for the two modes case, with $\bar{n}_{\text{c}}$ and $\bar{n}_{\text{s}}$ respectively the average phonon number in the COM and stretch mode.}
\label{fig:fidelityvstz}
\end{figure}

In Fig.~\ref{fig:fidelityvstz} we show the process fidelity of the proposed gate for $\delta/2\pi = 1$\,kHz in the single mode approximation with two different $\bar n$ along with a more thorough calculation including the stretch mode. With the single mode approximation, shown in solid blue and dashed orange lines, $\bar F$ exceeds 99\% at relatively low tweezer strength, $\omega_{\text{tw}}/\omega_{\text{com}}\gtrsim 0.1$. Fidelities higher than 99.9\% can be obtained for $\omega_{\text{tw}}/\omega_{\text{com}} \gtrsim 0.21$.
Higher tweezers intensity also allow us to perform faster gates at larger detunings while maintaining high fidelities. The green pentagons show the process fidelity including the contribution of the stretch mode and confirm the validity of the initial single mode approximation. When considering the contribution from both modes, we need to take into account a small correction to the electric field frequency $\mu$. This originates from the perturbation induced by the presence of tweezers on the original eigenmodes of the system. The deviation can be calculated either by perturbation theory (two ions) or by exact diagonalization (four ions). The corrections for the four ion crystal are shown in the second row of Table.~\ref{Table_4ions}. 

Finally, we study the scalability of the proposed scheme. We consider four ions in a harmonic potential under the full tweezer Hamiltonian and including all four modes of motion. Similar to the two mode calculation above, we correct $\delta$ given the COM mode shift. This correction depends on which ion pair the gate is implemented on and is calculated by exact diagonalization. The process fidelity for each ion pair at $\omega_\text{tw}=2\pi\times 254$\,kHz, as reported in Tab.\ref{Table_4ions}, does not degrade compared to the two ion crystal. This demonstrates the viability of extending this scheme to larger ion crystals, subject to laser power limitations because the required tweezer intensity scales as $\propto N$. Note, however, that since the gate does not require the Lamb-Dicke regime, the required tweezer power can be limited by lowering $\omega_\text{com}$ considerably.


\begin{table}[h]
\centering
\caption{Fidelities and detunings for all combinations of pairs in a four ion chain. All modes are in the ground state of motion. The tweezer strength is taken to be $\omega_{\text{tw}} = 2\pi\times 257$\,kHz and 
$\omega_{\text{com}} = 2\pi\times 1 $\,MHz.}
\label{Table_4ions}
\resizebox{\columnwidth}{!}{%
\begin{tabular}{c|c|c|c|c}
Pair & \begin{tikzpicture}\filldraw[fill=blue!40!white, draw=black] (0,0) circle (.15cm)node {$1$};\filldraw[fill=blue!40!white, draw=black] (.33,0) circle (.15cm)node {$2$};\draw (.66,0) circle (.1cm);\draw (.99,0) circle (.1cm);\end{tikzpicture} & \begin{tikzpicture}\filldraw[fill=blue!40!white, draw=black] (0,0) circle (.15cm)node {$1$};\draw (.33,0) circle (.1cm);\filldraw[fill=blue!40!white, draw=black] (.66,0) circle (.15cm)node {$3$};\draw (.99,0) circle (.1cm);\end{tikzpicture} & \begin{tikzpicture}\filldraw[fill=blue!40!white, draw=black] (0,0) circle (.15cm)node {$1$};\draw (.33,0) circle (.1cm);\draw (.66,0) circle (.1cm);\filldraw[fill=blue!40!white, draw=black] (.99,0) circle (.15cm)node {$4$};\end{tikzpicture} &\begin{tikzpicture}\draw (0,0) circle (.1cm);\filldraw[fill=blue!40!white, draw=black] (.33,0) circle (.15cm)node {$2$};\filldraw[fill=blue!40!white, draw=black] (.66,0) circle (.15cm)node {$3$};\draw (.99,0) circle (.1cm);\end{tikzpicture}  \\
\hline
$(1-F)\num{e4}$ \  & $3.7$   & $4.7$   &  $2.4$  &  $1.1$  \\

$(\omega_\text{com}-\mu) $ [kHz] \  & $1.212$   & $1.325$   &  $1.488$  &  $1.162$ 

\end{tabular}%
}

\end{table}

\noindent \emph{ Experimental considerations. } The tweezer potential takes the form: $\Phi_{|j\rangle}(\mathbf{r})\propto \alpha_{|j\rangle}(\lambda_\text{tw}) I(\mathbf{r})$ with $\alpha_{|j\rangle}(\lambda_\text{tw})$ the dynamic polarizability at the tweezer wavelength $\lambda_\text{tw}$ of qubit state $|j\rangle$ and $I(\mathbf{r})$ the intensity pattern~\cite{Grimm:2000}. Expanding a Gaussian intensity pattern with waist $w_0\gg l_m$ with $l_\text{m} = (2M\omega_\text{m})^{-1/2}$ we obtain $\Phi_{|j\rangle}(z)\approx \Phi_{|j\rangle}(0)+M\omega_{|j\rangle}^2 z^2/2$, with $\omega_{|j\rangle}^2 =-4 \Phi_{|j\rangle}(0)/(M w_0^2)$~\cite{Grimm:2000}. Here, we assumed that the tweezer has the largest curvature in the $z$-direction, and disregard the other directions. 

We have to identify qubit states with opposite dynamical polarizabilities such that $\omega_{|1\rangle}^2=-\omega_{|0\rangle}^2$. A convenient option are qubits encoded in the ground $S_{1/2}$ and metastable $D_{5/2}$ states of Ca$^+$, Sr$^+$ or Ba$^+$. The differential polarizabilities of these states can be tuned over a wide range by choosing the tweezer wavelength and Zeeman substate $m_j$ of the $D_{5/2}$ manifold~\cite{Kaur:2015}. 

Furthermore, it is beneficial to have no residual differential Stark shift at the center of the tweezer as this may lead to dephasing of the qubits in case of laser intensity fluctuations. The spin echo sequence will eliminate shot-to-shot variations, but not fluctuations within a single implementation. Vanishing differential Stark shift in the center of the tweezer can be straightforwardly obtained using non-Gaussian hollow tweezers~\cite{Schmiegelow:2016,Drechsler:2021}. Another solution is to use bichromatic tweezers with wavelengths $\lambda_\text{tw}^{(1)}$ and $\lambda_\text{tw}^{(2)}$ and beamwaists $w_1$ and $w_2$. We then require that in the center of the tweezer ($z = 0$): $\Phi_{|0\rangle}^{(1)}+\Phi_{|0\rangle}^{(2)}= \Phi_{|1\rangle}^{(1)}+\Phi_{|1\rangle}^{(2)}$ and $\Phi^{(1)}_{|0\rangle}/w_1^2+\Phi^{(2)}_{|0\rangle}/w_2^2=-\Phi^{(1)}_{|1\rangle}/w_1^2-\Phi^{(2)}_{|1\rangle}/w_2^2$. In the experimentally convenient limit where $w_1\ll w_2$, this reduces to: $\Phi_{|0\rangle}^{(1)}=-\Phi_{|1\rangle}^{(1)}$ and $\Phi_{|0\rangle}^{(2)}-\Phi_{|1\rangle}^{(2)}=2\Phi_{|1\rangle}^{(1)}$. Note that the frequency sum and difference should not be close to any transition as this will lead to additional Stark shifts or photon scattering.

As a practical example, we consider the qubit states $|0\rangle = |S_{1/2},m_j=1/2\rangle$ and $|1\rangle = |D_{5/2},m_j=3/2\rangle$ in $^{40}$Ca$^+$~\cite{Schindler:2013}. We obtain $\Phi_{|0\rangle}^{(1)}=-\Phi_{|1\rangle}^{(1)}$ at around $\lambda_1=770$\,nm~\cite{Kaur:2015}. The second requirement can be met by setting $\lambda_2\approx 900$\,nm. The relative close proximity of the $D_{5/2}\rightarrow P_{3/2}$ transition at 854\,nm causes photon scattering $\Gamma_\text{sc}$ which limits the attainable coupling strength. We estimate $\Gamma_\text{sc}=\Phi_{|j\rangle}^{(i)}\Gamma_\text{tr}/\Delta^{(i)}_{|j\rangle,\xi}$ for each transition $\xi$, state $|j\rangle$ and tweezer $i$ with $\Gamma_\xi$ the transition linewidth and $\Delta^{(i)}_{|j\rangle,\xi}$ the frequency detuning. Demanding $\Gamma_\text{sc}/2\pi\lesssim 1$\,s$^{-1}$, we find $|\Phi^{(i)}_{|j\rangle}|\lesssim$~20\,MHz for all $i$ and $|j\rangle$. This results in $|\omega_\text{tw}|\lesssim 2\pi\times$~70\,kHz for $w_2\gg w_1=$~1\,$\mu$m.

It is also possible to use qubits that are encoded in the ground $S_{1/2}$  hyperfine or Zeeman states of the ions. However, it is much harder to obtain sizeable differential Stark shifts between such states~\cite{Grimm:2000,Lee:2005}. One solution is to make use of quadrupole transitions~\cite{Haffner:2003a,Aolita:2007a}.  These have coupling strengths that are typically $\sim2\pi a_0/\lambda\sim 10^{-3}-10^{-4}$ times smaller than for dipole allowed transitions, with $a_0$ the Bohr radius, but have highly suppressed photon scattering rates even at small detunings. Tuning the tweezer wavelength far away from all dipole-allowed transitions, the differential Stark shift originates from the quadrupole transitions alone~\cite{Grimm:2000,Lee:2005,Haffner:2003a}. In case only a single transition $k'$ obeys $\Delta_{k'}\ll \omega_{0}$, with $\omega_0$ the frequency difference between the qubit states and $\Delta_{k'}$ the detuning, we can make a two-level approximation for the transition $|0\rangle\rightarrow |k'\rangle$ and obtain: 
$\Phi_{|0\rangle}\approx\nu_\text{Dipole}+\Omega_{k'}^2/(4\Delta_{k'})$ while $
\Phi_{|1\rangle}\approx \nu_\text{Dipole}$ with $\Omega_{k'}$ the Rabi frequency.
Approximating the Stark shift due to the dipole transitions to arise mainly from a single (effective) transition, $\nu_\text{Dipole}=\Omega_\text{Dipole}^2/(4\Delta_\text{Dipole})$, we get $\Omega_{k'}^2/(4\Delta_{k'})=-\nu_\text{Dipole}$ if we set $\Delta_{k'}=-\epsilon^2\Delta_\text{Dipole}$ with $\Omega_{k'}=\epsilon\times \Omega_\text{Dipole}$. The detuning $\Delta_\text{Dipole}$ can be estimated as the frequency difference between the quadrupole transition and the strong D1 and D2 transitions and lies typically in the 100-THz range~\cite{NIST}. Therefore, we require $\Delta_{k'}\sim~1-100$\,MHz for $\epsilon=10^{-3}-10^{-4}$. Since we require in addition  $\Delta_{k'}<\Omega_{k'}$ to avoid driving the transition, we get differential Stark shifts of $\sim~0.1-10$\,MHz and $\omega_\text{tw}\sim 2\pi (30-300)/\sqrt{M_u}$\,kHz, with $M_u$ the mass of the ion in atomic mass units and $w_0=$~1\,$\mu$m. By comparison, switching to a Laguerre-Gaussian mode with radial index $p=1$, the Rabi frequency in the center of the tweezer vanishes, whereas $\omega_\text{tw}$ remains unaltered for the same $w_0$. In this situation, we only require $\Delta_{k'}<\Omega_{k'}(z_\text{max})$, with $z_\text{max}$ the maximum amplitude of motion of the ions during the gate. For the presented calculations $z_\text{max}\sim$~10\,nm~$\ll w_0$ such that $\omega_\text{tw}$ can be significantly larger than for Gaussian tweezers. 

\emph{Discussion. } We have proposed and analysed an architecture for performing quantum computation with trapped ions that is based on optical tweezers in combination with oscillating electric fields. The infrastructural simplicity of the latter makes the scheme attractive, while the addressed tweezers at the same time allow for individual addressing of the ions and therefore universality. The scheme does not rely on the Lamb-Dicke approximation and it is independent of the qubits separation, as the electric field couples only to the COM mode of the ion crystal. Residual qubit-phonon entanglement that may lead to decoherence is prevented by a spin-echo sequence. Taking experimental considerations into account, the scheme can be performed on optical qubits. For ground state qubits hollow tweezers such as those derived from e.g.\ Laguerre-Gaussian modes~\cite{Schmiegelow:2016,Drechsler:2021} would be the preferred choice. The challenge will be to supply sufficient curvature to such tweezers while maintaining excellent control. For this it would seem necessary to actively stabilize the power and to point in the tweezers, for instance, by regularly performing service measurements and performing feedback by e.g. spatial light modulators. Finally, it seems feasible to consider a fast gate version of the proposed gate, in analogy to Ref.~\cite{Vogel:2019} where electric field pulses are combined with Rydberg excitation of the trapped ions in order to implement quantum logic gates.

\section*{Acknowledgements}
We gratefully acknowledge discussions with Robert Spreeuw. This work was supported by the Netherlands Organization for Scientific Research (Grant Nos. 680.91.120 and 680.92.18.05, R.G., M.M. and R.X.S.). A.S.N is supported by the Dutch Research Council (NWO/OCW), as part of the Quantum Software Consortium programme (project number 024.003.037).




%

\end{document}